\documentclass{jpp}
\expandafter\let\csname equation*\endcsname\relax
\expandafter\let\csname endequation*\endcsname\relax
\usepackage{amsmath}
\usepackage{amsfonts}
\usepackage{amssymb}
\usepackage{bbold}
\usepackage{url}
\usepackage{graphicx}
\usepackage{setspace}
\usepackage{color}
\begin{document}
\title{Bayesian approach for validation of runaway electron simulations}
\author{A.E. Järvinen\aff{1, 2}\corresp{\email{aaro.jarvinen@vtt.fi}}, T. Fülöp\aff{3}, E. Hirvijoki\aff{4}, M. Hoppe\aff{5}, A. Kit\aff{2}, and J. Åström\aff{6, 2}}

\affiliation{\aff{1}VTT Technical Research Centre of Finland, FI-02044 VTT, Finland
\aff{2}University of Helsinki, FI-00014 Helsinki, Finland
\aff{3}Chalmers University of Technology, SE-412 96 Göteborg, Sweden
\aff{4}Aalto University, FI-00076 AALTO, Finland
\aff{5}Ecole Polytechnique Fédérale de Lausanne (EPFL), Swiss Plasma Center, CH-1015 Lausanne, Switzerland
\aff{6}CSC-IT Center for Science, FI-02101 Espoo, Finland}

\maketitle

\begin{abstract}
  Plasma-terminating disruptions in future fusion reactors may result in conversion of the initial current to a relativistic runaway electron beam.
  Validated predictive tools are required to optimize the scenarios and mitigation actuators to avoid the excessive damage that can be caused by such events.
  Many of the simulation tools applied in fusion energy research require the user to specify several input parameters that are not constrained by the available experimental information.
  Hence, a typical validation exercise requires multiparameter optimization to calibrate the uncertain input parameters for the best possible representation of the investigated physical system.
 The conventional approach, where an expert modeler conducts the parameter calibration based on domain knowledge, is prone to lead to an intractable validation challenge.
  For a typical simulation, conducting exhaustive multiparameter investigations manually to ensure a globally optimal solution and to rigorously quantify the uncertainties is an unattainable task, typically covered only partially and unsystematically.
  Bayesian inference algorithms offer a promising alternative approach that naturally includes uncertainty quantification and is less subjective to user bias in choosing the input parameters.
  The main challenge in using these methods is the computational cost of simulating enough samples to construct the posterior distributions for the uncertain input parameters.
  This challenge can be overcome by combining probabilistic surrogate modelling, such as Gaussian Process regression, with Bayesian optimization, which can reduce the number of required simulations by several orders of magnitude.
  Here, we implement this type of Bayesian optimization framework for a model for analysis of disruption runaway electrons, and explore for simulations of current quench in a JET plasma discharge with an argon induced disruption.
We use this proof-of-principle framework to explore the optimum input parameters with uncertainties in optimization tasks ranging from one to seven dimensions.
\newline
\newline
The relevant Python codes that are used in the analysis are available via \break \url{https://github.com/aejarvin/BO_FOR_RE_SIMULATIONS/} 
\end{abstract}

\textbf{Keywords: }
Bayesian inference, Bayesian optimization, DREAM, Runaway electrons, Fusion energy, Uncertainty quantification

\section{Introduction}
Runaway electrons (RE) pose one of the leading concerns regarding the integrity and duty cycle of future fusion reactors \citep{BoozerNF2018}. As RE generation in disruptions is exponentially sensitive to the plasma current, $I_\text{p}$ \cite[]{RosenbluthNF1997}, and $I_\text{p}$ is projected to increase from the level of a few MA in large present-day tokamaks to the range of 10 -- 20 MA in power plant relevant tokamaks, an unmitigated RE beam at reactor-scale would be expected to cause severe damage, extended downtime, and repair costs \cite[]{BoozerNF2018, BreizmanNF2019}. Therefore, there is a strong need for validated tools to predict and avoid disruptions and RE beam generations when entering the era of reactor-scale devices. 

To address this need, several numerical tools have been developed for disruption and RE analysis, such as the nonlinear magnetohydrodynamic (MHD) codes JOREK \cite[]{HuysmansNF2007}, and NIMROD \cite[]{SovinecJCP2004}, the kinetic code CQL3D \cite[]{HarveyPoP2000}, and the fluid-kinetic framework DREAM \cite[]{HoppeCPP2021}. However, validating these simulation tools with experimental data is complicated, as typically some of the input parameters of the simulations are not well constrained by the available experimental information. In such a situation, 
the user must first specify values for these uncertain input parameters to calibrate the model to appropriately represent the investigated system. This challenge is common to other numerical tools applied in magnetic confinement fusion research as well, such as scrape-off layer plasma simulations conducted with SOLPS-ITER \cite[]{WiesenJNM2015}.  A conventional approach to calibrate the model is to use previous experience and domain knowledge to conduct the necessary parameter fitting manually, aiming to find a set of input parameters that minimizes the discrepancy between the synthetic and measured diagnostic data. The remaining discrepancy is then evaluated and documented as a degree of validity of the model. 

However, this type of expert modeller approach becomes intractable as the number of uncertain parameters increases. With multiple uncertain input parameters, manual optimization is prone to lead to subjective reasoning for the trajectory selection through the space of optimized parameters. As a result, the degree of confidence on the obtained solution and its uncertainty is likely to remain ambiguous. Furthermore, the ad hoc nature of subjective reasoning is prone to lack of scientific rigour, which is problematic from the point of view of aiming to establish objective and reproducible scientific results. These challenges can be alleviated by applying a regimented approach, such as grid search, instead of subjective reasoning, but such an approach is intractable to conduct manually with multiple uncertain input parameters and is computationally inefficient when operating with costly simulations. Due to these inefficiencies of the expert modeller approach, an optimization algorithm that would take the human out-of-the-loop, provide a rigorous systematic approach with natural uncertainty quantification, and select the samples from the search space in a computationally efficient manner would be a very attractive alternative approach \cite[]{brochu2010tutorial, ShahriariIEEE2016, frazier2018tutorial}. 

In this study, approximate Bayesian computation (ABC) and Bayesian optimization (BO) are applied to find the optimal values and provide inverse uncertainty quantification of uncertain input parameters  \cite[]{brochu2010tutorial, ShahriariIEEE2016, frazier2018tutorial,  MarinCS2012} in DREAM runaway electron simulations.  Inverse uncertainty quantification here refers to quantifying the uncertainty of the uncertain input parameters of the model given observed experimental data \cite[]{WuNuclEngDes2018, OberkampfProgAeroSci2002}. The aim of this study is to provide a proof-of-principle approach to using these methods in calibrating the uncertain input parameters in RE simulations, while the methodology could be used broadly in validating other predictive tools within magnetic confinement fusion research. The implementation is based on the Bayesian Optimization for Likelihood-Free Inference (BOLFI) method of the Engine for Likelihood-Free Inference (ELFI) Python software package \cite[]{BOLFI, ELFI}.

\section{Bayesian approach}
This Section describes the methodology used in this study. Subsection \ref{subsec:ABC} provides an overview of Approximate Bayesian Computation (ABC), Subsection \ref{subsec:BO} an introduction to Bayesian Optimization (BO), Subsection \ref{subsec:GPR} describes the usage of Gaussian Process Regression as a probabilistic surrogate model in BO, and Subsection \ref{subsec:ACQ} describes the functionality of the acquisition functions in BO.

\subsection{Approximate Bayesian computation}\label{subsec:ABC} 

Bayesian inference aims to establish the conditional probability distribution, $P(\mathbf{x}|D)$, called the posterior, of the uncertain input parameters, $\mathbf{x}$, given observed experimental measurements, $D$. $P(\mathbf{x}|D)$ represents the best estimate and uncertainty of the input parameters for the investigated system. Bayesian inference applies the Bayes' theorem: 
 \begin{equation}
     P(\mathbf{x}|D) = \frac{P(D|\mathbf{x})P(\mathbf{x})}{P(D)},
 \end{equation}
 which states that the posterior is proportional to the likelihood of $D$ given $\mathbf{x}$, $P(D|\mathbf{x})$, multiplied by the prior probability distribution for $\mathbf{x}$, $P(\mathbf{x})$. The marginal probability of the experimental measurements, $P(D)$, represents an integral over all possible data generating input values, $\int P(D|\mathbf{x}')P(\mathbf{x}')d\mathbf{x}'$, which would be computationally challenging to evaluate. However, it is typically sufficient to establish the relative posterior probabilities of various values of $\mathbf{x}$. Therefore, $P(D)$ does not need to be directly evaluated, and it is sufficient to apply the proportionality: 
 \begin{equation}
     P(\mathbf{x}|D) \propto P(D|\mathbf{x})P(\mathbf{x}). 
 \end{equation}
 
 When the likelihood function, $P(D|\mathbf{x})$, is either not available analytically or cannot be evaluated within the available computational or time resources, the standard alternative is to use approximate Bayesian computation (ABC) \cite[]{MarinCS2012, JarvenpaaAAS2018}. ABC aims to establish the approximate Bayesian posterior:
 \begin{equation}\label{ABC}
     P_\text{ABC}(\mathbf{x}|D) \propto P(\mathbf{x}) \int H(\epsilon - \Delta (D, y)) P(y|\mathbf{x})dy,
 \end{equation}
where $y \in \mathbb{R}^d$ is data generated with the simulation model with input parameters $\mathbf{x}$, and $\Delta: \mathbb{R}^d\times\mathbb{R}^d \rightarrow \mathbb{R}_{>0}$ is a discrepancy function between the simulated and measured data. $\epsilon$ represents the threshold parameter controlling the trade-off between posterior estimation accuracy and efficiency, and $H(x)$ is the Heaviside step function which takes a value 1 whenever $\epsilon$ is greater than the discrepancy. Small values of $\epsilon$ lead to more accurate approximate posteriors, but also increase the computational challenge. 

One of the simplest ABC algorithm that could be applied to numerically estimate the integral (Eq. \ref{ABC}) is rejection sampling \cite[]{MarinCS2012, JarvenpaaAAS2018, LintusaariSysBio2017}: 
\begin{enumerate}
    \item Draw a random sample from $\mathbf{x}' \sim P(\mathbf{x})$.
    \item Generate $y' \sim P(y|\mathbf{x}')$.
    \item Accept $y'$ if $\Delta (D, y') \le \epsilon$.
    \item Go back to step (i) until sufficient number of accepted samples are collected.
\end{enumerate} 
The accepted values represent the approximate posterior distribution. The drawback of the standard rejection sampling is the number of required function evaluations. For a typical simulation tool in magnetic confinement fusion, a function evaluation takes at least several minutes and more typically hours or days. Therefore, it is not computationally feasible to collect sufficiently many samples to get an accurate ABC posterior distribution using a rejection sampler. Furthermore, a rejection sampler with a small threshold parameter, $\epsilon$, is computationally very inefficient as a large fraction of the sampled function evaluations are rejected. While it is possible to improve the efficiency by applying approaches such as Markov Chain Monte Carlo (MCMC) \cite[]{MarjoramPNAS2003}, the direct sampling based ABC algorithms are still expected to be computationally too costly for the type of applications targeted in this work.  

The inefficiency of the direct sampling based ABC approach can be circumvented by applying Bayesian optimization to traverse the space of optimized input parameters  \cite[]{BOLFI, Jarvenpaa2019}. This leads to a probabilistic surrogate model based ABC approach that is observed to be several orders of magnitude more efficient in terms of full function evaluations than the direct sampling based ABC algorithms. At each step of the algorithm, the ABC posterior is estimated using the surrogate model as $P_\text{ABC}(\mathbf{x}|D) \propto P(\mathbf{x}) \mathbb{P}(\Delta_\mathbf{x} < \epsilon)$, where the probability is computed using the probabilistic surrogate model.

\subsection{Bayesian optimization} \label{subsec:BO}

Bayesian optimization (BO) offers a powerful approach for global optimization of costly-to-evaluate, non-convex functions, without access to first- or second-order derivatives  \cite[]{brochu2010tutorial, ShahriariIEEE2016, frazier2018tutorial}. The problem of finding optimal values, $\mathbf{x_*}$, for the uncertain input parameters, $\mathbf{x}$, can be represented as a task of finding the optimum of a non-linear function $f(\mathbf{x})$ of a compact set $\mathcal{A}$, called search space in this report. If $f(\mathbf{x})$ represents the discrepancy between the synthetic and measured diagnostic data, then the problem can be formulated as:
\begin{equation}
    \mathbf{x_*} = \arg \min_{\mathbf{x} \in \mathcal{A}} f(\mathbf{x}). 
\end{equation}
 
 The target for a BO algorithm is to be able to traverse the search space efficiently in terms of function evaluations and to find the globally optimum solution by applying prior belief about the optimized function and by balancing exploration and exploitation of the search space \cite[]{brochu2010tutorial, ShahriariIEEE2016, frazier2018tutorial}. In exploitation, samples are collected in regions of the search space that are known to lead to near optimal function values based on prior belief, and in exploration, samples are collected in regions that encompass a large uncertainty.

 A standard BO algorithm consists of two main components \cite[]{brochu2010tutorial, ShahriariIEEE2016, frazier2018tutorial}: 
 \begin{enumerate}
     \item A probabilistic model for the objective function. 
     \item An acquisition function for recommending the next sampling point.
 \end{enumerate}
The probabilistic model represents essentially a low evaluation cost surrogate model for the  objective function, and the uncertainties retained in the probabilistic model represent the degree of confidence on the surrogate model predictions. The acquisition function applies the mean and variance of the probabilistic model to balance exploitation and exploration. The collected sample values are then used to update the probabilistic surrogate model.    

\subsection{Gaussian process regression}\label{subsec:GPR}

The usual choice for the probabilistic model is to use Gaussian Process Regression (GPR), also known as Kriging \cite[and references therein]{Stein1999, RW2006}. Kriging surrogate-based optimization was previously used for parameter optimization of plasma transport codes by \cite{RodriguezFED2018}. Other examples of GPR applications in plasma physics can be found in \cite[and references therein]{HoNF2019, ChilenskiNF2017, ChilenskiNF2015, vonNessiPoP2012, LiRSI2013, RomeroNF2013, vonNessiRSI2013}.  

GPR is a Bayesian regression technique and is very powerful for interpolating small sets of data as well as retaining information about the uncertainty of the regression \cite[and references therein]{Stein1999, RW2006}. Gaussian process (GP) is a stochastic process, for which any finite collection of random values has a multivariate normal distribution. GP is specified by the mean function, $m(\mathbf{x}) = \mathbb{E}[f(\mathbf{x})]$, and the covariance function, $k(\mathbf{x},\mathbf{x'}) = \mathbb{E}[(f(\mathbf{x}) - m(\mathbf{x}))(f(\mathbf{x'}) - m(\mathbf{x}'))]$ \cite[]{RW2006}: 
\begin{equation}
    f(\mathbf{x}) \sim \mathcal{GP}(m(\mathbf{x}), k(\mathbf{x},\mathbf{x'})).
\end{equation}
The GPR represents a family of functions, for which the point to point variance is described by the covariance function.
Usually, the mean function is assumed as zero, although other assumptions are possible. This means that the mean of the prior assumption on variation of the objective function value when propagating from a collected data point is zero. The covariance function or kernel describes the smoothness assumption on the possible functions $f$. GP essentially describes a normal distribution over possible functions, $f$, conditioned with observations \{$(\mathbf{x}_i, f_i)$, $i = 1,...,n$\}. Assuming a collection of possibly noisy observations with a Gaussian noise variance $\sigma_n^2$, the posterior probability distribution function of $f$ at point $\mathbf{x}$ is Gaussian with posterior mean $\mu_n(\mathbf{x})$ and posterior variance $v_t(\mathbf{x}) + \sigma_n^2$:
\begin{equation}
    f(\mathbf{x}) \sim \mathcal{N}(\mu_n(\mathbf{x}), v_t(\mathbf{x}) + \sigma_n^2).
\end{equation}
Assuming a zero mean function, $m(\mathbf{x})$, the mean and variance can be obtained as 
\begin{equation}
         \mu_n(\mathbf{x_*}) = \mathbf{k}_*^\intercal\mathbf{K}_n^{-1}\mathbf{f_n},
\end{equation}
\begin{equation}
    v_n(\mathbf{x_*}) = k(\mathbf{x_*},\mathbf{x_*}) - \mathbf{k}_*^\intercal\mathbf{K}_n^{-1}\mathbf{k}_*,
\end{equation}
where $\mathbf{k_*}$ represents the vector of covariances between the test point, $\mathbf{x_*}$, and the $n$ observations, $f_n$ is a vector of the $n$ observations, and $\mathbf{K}$ is the covariance matrix \cite[]{ BOLFI, RW2006}.
Since the function evaluations in this work are deterministic, the $\sigma_n$ term is constrained to a low value that does not impact the predictions. An estimate of the likelihood at $w$ is given by
\begin{equation}\label{posterior}
    \mathbb{P}(\Delta_w < \epsilon) = F\left(\frac{\epsilon - \mu_n(w)}{\sqrt{v_n(w) + \sigma_n^2}}\right), 
\end{equation}
where $F(x)$ is the cumulative distribution function of $\mathcal{N}(0,1)$. In this work, we will set $\epsilon$ equal to the current optimal value provided by the probabilistic surrogate model, which is also the default in BOLFI \cite[]{BOLFI}.

A key step in building a GP regression is to select the covariance function or kernel \cite[]{RW2006}. Usually the default choice is the radial basis function (RBF), also known as squared exponential or Gaussian kernel: 
\begin{equation}
    k_\text{RBF}(x_i, x_j) = \sigma_f^2 \text{exp}\left(-\sum_{i=1}^{d}\frac{\left(x_{i,k} - x_{j,k}\right)^2}{2l_k^2}\right),
\end{equation}
where $\mathbf{l} = [l_1,...,l_d]$ is a vector of covariance lengthscales for each dimension, $d$, and $\sigma_f^2$ is the variance. In the applications in this work, the single constant $\mathbf{l}$ was observed to be often too restrictive and a rational quadratic kernel (RQ) was used instead: 
\begin{equation}
    k_\text{RQ}(x_i, x_j) = \sigma_f^2 \left(1 + \sum_{i=1}^{d}\frac{\left(x_{i,k} - x_{j,k}\right)^2}{2\alpha l_k^2}\right)^{-\alpha},
\end{equation}
which is equivalent to summing many RBF kernels with varying $\mathbf{l}$. The hyperparameter $\alpha$ represents the relative weighting between large and small $\mathbf{l}$ values. Before applying the model, the hyperparameters ($\mathbf{l}$, $\alpha$, $\sigma_f^2$, $\sigma_n$) must be optimized first. These can be estimated by maximizing the marginal log-likelihood \cite[]{RW2006}. The GPR library used in this work as well as in BOLFI is the Python GP framework GPy \cite[]{gpy2014}, which encompasses the applied optimization routines. 

\subsection{Acquisition function} \label{subsec:ACQ}

The acquisition function applies the mean and variance of the probabilistic surrogate model to recommend the next sampling point for the objective function \cite[]{brochu2010tutorial, ShahriariIEEE2016, frazier2018tutorial}. The acquisition functions are typically constructed to recommend sampling points that either encompass optimal predicted mean for the objective function, exploitation, or large uncertainty, exploration. The sampling point is selected by optimizing the acquisition function. Several acquisition functions have been developed and can be found in the reviews in \cite[]{brochu2010tutorial, ShahriariIEEE2016, frazier2018tutorial}. The acquisition function used in sequential sampling in this work is the lower confidence bound selection criterion (LCBSC), which is also the default acquisition function in BOLFI \cite[and references therein]{ brochu2010tutorial,BOLFI, SrinivasICML2010}. This function can be written as 
\begin{equation}\label{acq}
    \mathcal{A}_n(\mathbf{x}) = \mu_n(\mathbf{x}) - \sqrt{\eta_n^2v_n(\mathbf{x})},
\end{equation}
\begin{equation}
    \eta_n^2 = 2\text{ln}\left(\frac{n^{2d+2}\pi^2}{3\epsilon_\eta}\right),
\end{equation}
where the coefficient $\eta_n$ is the trade-off parameter between exploration and exploitation, $\epsilon_\eta \in (0,1)$ is a constant chosen by the user, and $d$ is the dimensionality of the search space. Optimizing the acquisition function provides a deterministic answer for the next sampling point. An example of the application of this acquisition function is shown in Section \ref{1D_search}.

Since the next sampling point is obtained deterministically for a given state of the surrogate model, the approach is naturally sequential: (1) the objective function is evaluated for the sampling location provided by the optimum of the acquisition function, (2) the GPR surrogate model is updated, and (3) the acquisition function is optimized again, using the updated GPR, to recommend the next sampling point. However, with complicated, multi-dimensional optimization tasks with computationally time consuming function evaluations, it would be more attractive to conduct several objective function evaluations in parallel to each other to reduce the overall time consumption of the optimization task, especially when suitable high-performance computing (HPC) resources are available. 

To conduct parallel BO, stochastic acquisition rules can be used. Various approaches for parallel BO and batch acquisition have been developed \cite[]{JarvenpaaAAS2018, Jarvenpaa2019, Thompson1933, KandasamyPMLR2018, Chapelle2011,  DBLP:journals/corr/abs-1903-09434}. In this work, the randmaxvar approach developed by \cite{Jarvenpaa2019} is used as the stochastic acquisition method. The approach is based on the maxvar acquisition rule also presented in \cite{Jarvenpaa2019}. The maxvar acquisition method recommends a sample in a location that encompasses the maximum variance of the unnormalized ABC posterior. Basically, due to the limited information, represented by the collected samples, there is uncertainty in the GPR representation of the objective function and this uncertainty is propagated as uncertainty of the unnormalized ABC posterior. The maxvar method aims to collect samples that lead to maximum reduction of this uncertainty. In the stochastic version of this method,  samples are collected from the distribution that represents the variance of the unnormalized ABC posterior. Since samples are collected stochastically, several samples can be collected without updating the GPR surrogate in between the samples, enabling parallelization. Furthermore, sampling can be done asynchronously by simply updating the GPR surrogate whenever new results are added to the dictionary of collected samples and by sampling new values whenever an idle processor becomes available.       


\section{Application to a JET runaway electron experiment}

This Section describes proof-of-principle applications of the methodology discussed in Section 2 for a RE experiment at JET. Subsection \ref{experiment} describes the investigated JET plasma discharge and the DREAM setup, Subsection \ref{1D_search} documents a 1D proof-of-principle optimization, Subsections \ref{5D_search},  \ref{7D_search}, and \ref{4D_6D} extend the search to 4 to 7 dimensional search spaces, and Subsection \ref{conv} evaluates the number of samples required for convergence as a function of dimensionality of the search. 

\subsection{Simulated experiment and DREAM setup} \label{experiment}
The BO and ABC approach discussed in the previous section is applied for DREAM runaway electron simulations of current quench (CQ) in the disruption of a JET discharge \#95135. This was a deuterium limiter plasma with an argon massive gas injection induced disruption, described in detail by \citet[]{ReuxPRL2021} and \citet[]{Brandstrom2021}.    DREAM is a numerical tool for self-consistently simulating the evolution of temperature, poloidal flux, and impurity densities, along with the generation and transport of REs in tokamak disruptions \citep[]{HoppeCPP2021}.
The DREAM simulations in this manuscript are similar to those presented by \citet[]{Brandstrom2021} with the exception that only a fluid model for RE electrons is used here, as kinetic simulations were not necessary for the proof-of-principle of the Bayesian approach.   For a full description of the physics model in DREAM, we refer the reader to Ref.~\citep[]{HoppeCPP2021}.

 The simulations are started at the peak of the total plasma current, $I_\text{p}$, obtained during the disruption.
 An instantaneous thermal quench (TQ) is assumed, after which all background plasma quantities, except the electron temperature, $T_\text{e}$, are evolved self-consistently. The post-disruption $T_\text{e}$ is instead given as an uncertain input parameter.
 The background plasma density is obtained from pre-disruption measurement with the high resolution Thomson scattering (HRTS) \citep{HRTS2004, HRTS2012}.
 Even though the uncertainty of the electron density measurement could be taken into account in the Bayesian approach, for simplicity we neglect it here. 

 The argon density in the plasma is obtained from the estimated amount of injected argon, volume of the vessel, and fraction of argon that is assimilated.
 Argon is assumed to be uniformly distributed in the plasma. While the injected amount can be obtained from the experiment, $N_\text{Ar} \sim 8\times10^{20}$ atoms \cite[]{Brandstrom2021}, the assimilated fraction is given as an uncertain input parameter, $f_\text{Ar}$.

 Hot-tail and Dreicer RE generation mechanisms were not self-consistently included in the simulations. Instead, a RE seed profile is given as an uncertain input to generate a RE beam through the avalanche mechanism.
  The initial total plasma current $I_{\text{p}}$ (combination of ohmic and prescribed runaway seed current) is constrained to match the experimentally measured peak value, $I_\text{p} \sim 1.42$ MA, by adjusting the initial electric field in order for the avalanche multiplication factor to to be constrained.
 The electric field is evolved self-consistently, during the CQ and RE plateau simulations.
 The conductivity of the wall is controlled by a characteristic wall
 time, $\tau_\text{wall} = L_\text{ext}/R_\text{wall}$, that
 is provided as an uncertain input parameter. Here, $L_\text{ext}$ is
 the external inductance and $R_\text{wall}$ the resistance of the
 wall.

 The full list of uncertain input parameters are $T_\text{e}$, $f_\text{Ar}$, RE seed distribution, and $\tau_\text{wall}$.
 In addition, the RE seed distribution is scaled with a multiplier such that the RE plateau current matches the experimentally measured value.
 This is done iteratively with a binary search algorithm. The algorithm is initialized by multiplying or dividing the multiplier by $10^{3}$ at each iteration until the predicted plateau current passes the experimentally measured value, after which binary search is applied to converge to the optimum. If the multiplier is reduced below a small value, $\sim 10^{-3}$, the search algorithm stops without finding a solution that matches the RE plateau current and the objective function value at the point of reaching that threshold is propagated through the algorithm. This is simply due to the fact that it cannot be assumed that for all input values within the search space it is possible to find a multiplier that would enable matching the RE plateau current.       


\begin{figure}
    \centering
    \includegraphics[width=0.65\textwidth]{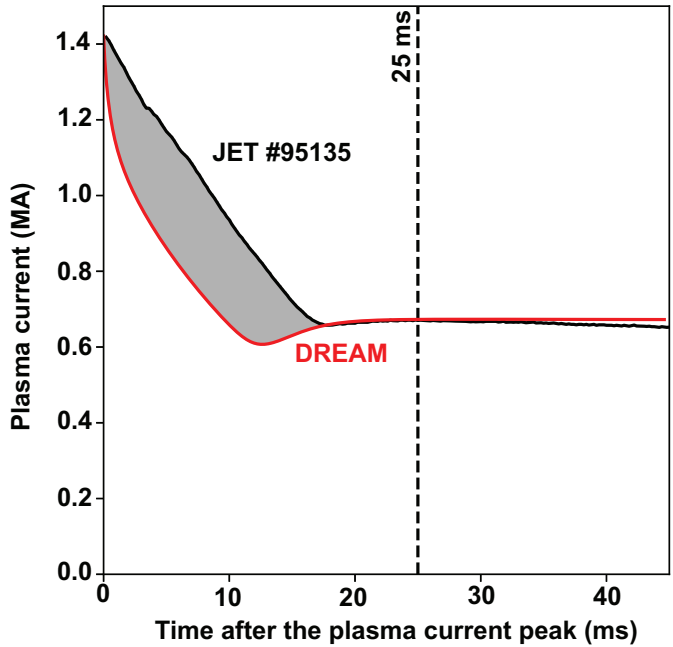}
    \caption{Illustration of the discrepancy between the measured (black) and simulated (red) plasma current for an example simulation. The grey area represents the L1-norm used as the discrepancy metric in this study. The vertical dashed line indicates the temporal extent of the application of the discrepancy function.}
    \label{fig:distance}
\end{figure}

In the proof-of-principle examples that follow, the objective function is chosen as the L1-norm of the discrepancy between the measured and predicted $I_\text{p}$ during the CQ (Fig. \ref{fig:distance}): 
\begin{equation}\label{objective}
    \Delta = \lVert I_\text{p}^\text{measured} - I_\text{p}^\text{DREAM}\rVert_1 .
\end{equation}
Effectively this discrepancy function calculates the area between the two curves in Figure 1. To avoid accumulating excessive integration within the RE plateau, the discrepancy is only calculated between 0 and 25 ms from the current peak.

\subsection{1D search space}\label{1D_search}

The first proof-of-principle test is to apply the Bayesian approach to search for the post TQ $T_\text{e}$ that minimises the discrepancy function (Eq. \ref{objective}). The other uncertain input parameters are fixed as $f_\text{Ar} = 15$\%, uniform RE seed profile, and $\tau_\text{wall} = 5$ ms. A rational quadratic kernel is used and the lengthscale is constrained to be below 1.0. This is done to avoid the model becoming overconfident in regions that have not been sampled yet. If the lengthscale converges to a large value, it can suppress exploration prematurely and prevent the algorithm from finding the optimal solution. More generally it seems that in BO algorithms it is probably better to have a model that has capability to overfit rather than underfit the data. The BO algorithm is interpolating solutions within the search space and an overfitting model will just encourage exploration while an underfitting model might not have the generalization capability to fit the solution near the optimum. The acquisition function is the LCBSC (Eq. \ref{acq}) with exploration constant  $\epsilon_\eta = 0.2$. The search space for $T_\text{e}$ is bounded between 1.0 and 20 eV. A uniform prior uncertainty distribution is assumed. 

\begin{figure}
    \centering
    \includegraphics[width=\textwidth]{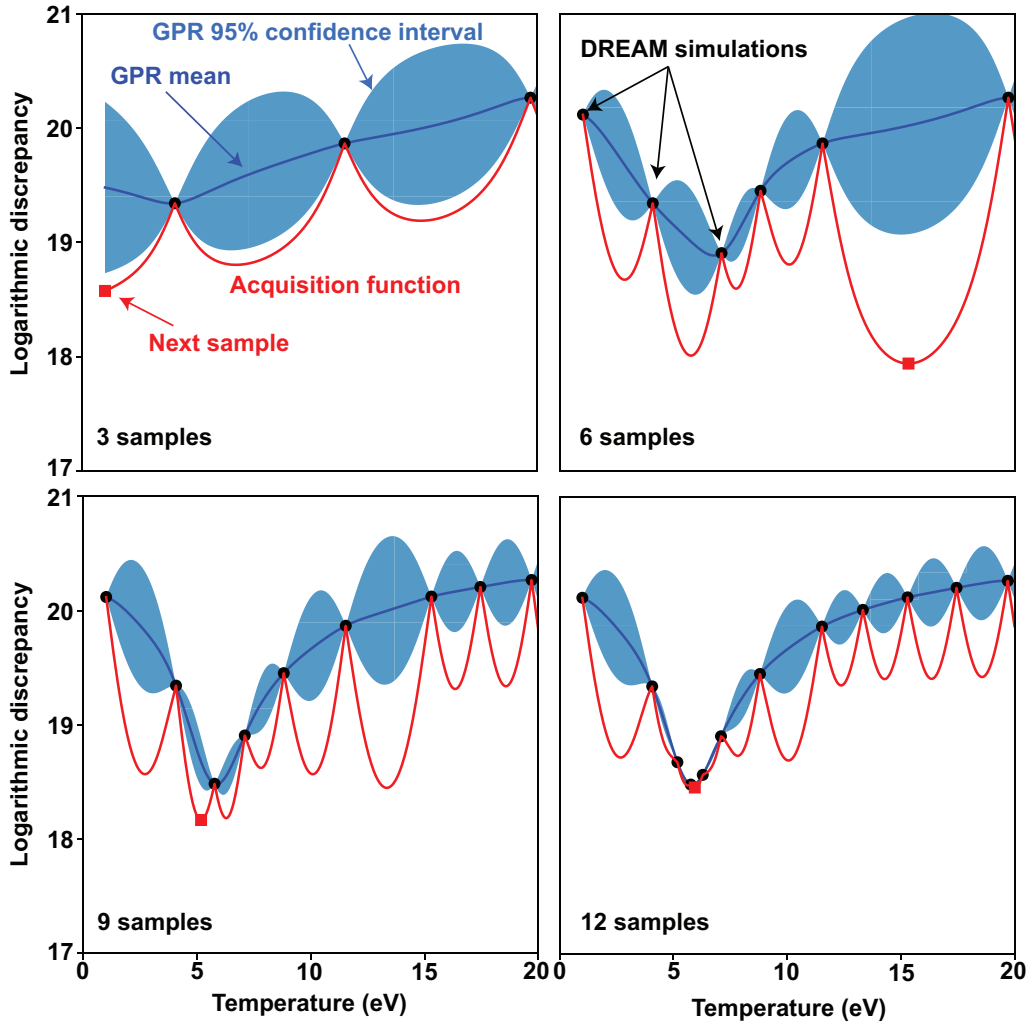}
    \caption{Illustration of the progress of the BO algorithm in the 1D example after 3, 6, 9, and 12 samples. The dark blue lines represent the mean and the light blue regions represent the 95\% confidence interval of the GPR. The red lines represent the acquisition function and the red squares illustrate the optimum of the acquisition function that provides the next sampling location. The black circles repsent the collected objective function values obtained through DREAM simulations.}
    \label{fig:12_samples}
\end{figure}

\begin{figure}
    \centering
    \includegraphics[width=\textwidth]{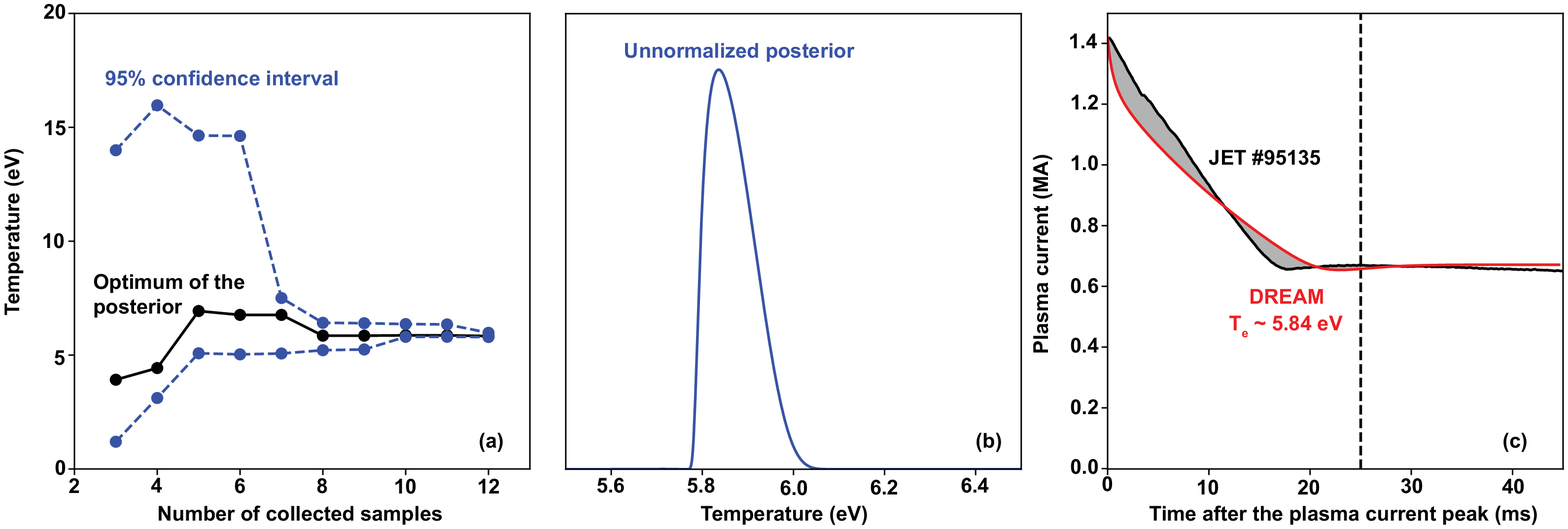}
    \caption{1D search space results. (a) Optimum of the posterior (black) and 95\% confidence interval (blue dashed) as a function of number of collected samples. (b) Unnormalized posterior distribution after collecting 12 samples. (c) Measured (black) and predicted (red) plasma current as a function of time. The predicted value is conducted at the optimum of the posterior after 12 collecting 12 samples.}
    \label{fig:1D_convergence}
\end{figure}

Within less than 10 iterations, the algorithm starts to converge to the optimum value (Fig. \ref{fig:12_samples}). The first 3 samples are collected randomly from the uniform prior distribution. After these are collected, the GPR surrogate model is fitted to the data (Fig. \ref{fig:12_samples}). The mean and variance of the GPR are applied by the acquisition function to recommend the next sampling location. The algorithm proceeds by choosing the location of the minimum value of the acquisition function. By proceeding like this, the algorithm converges near the optimum value with a narrow confidence interval around the sample number 7 to 8 (Fig. \ref{fig:1D_convergence}a). Since the prior distribution is uniform, the posterior probability distribution can be obtained from the GPR by applying Eq. \ref{posterior} (Fig.~\ref{fig:1D_convergence}b). Finally, the simulated current value with the temperature value providing the highest posterior probability, $T_\text{e} \sim 5.84$ eV, can be compared to the experimental measurements (Fig.~\ref{fig:1D_convergence}c). The results indicate that by only adjusting the constant post thermal quench $T_\text{e}$, the model is not able to reproduce the experimentally measured CQ rate. Alternatively, these results indicate that the background plasma resistivity is changing during the CQ.

\subsection{5D search space}\label{5D_search}

\begin{figure}
    \centering
    \includegraphics[width=0.95\textwidth]{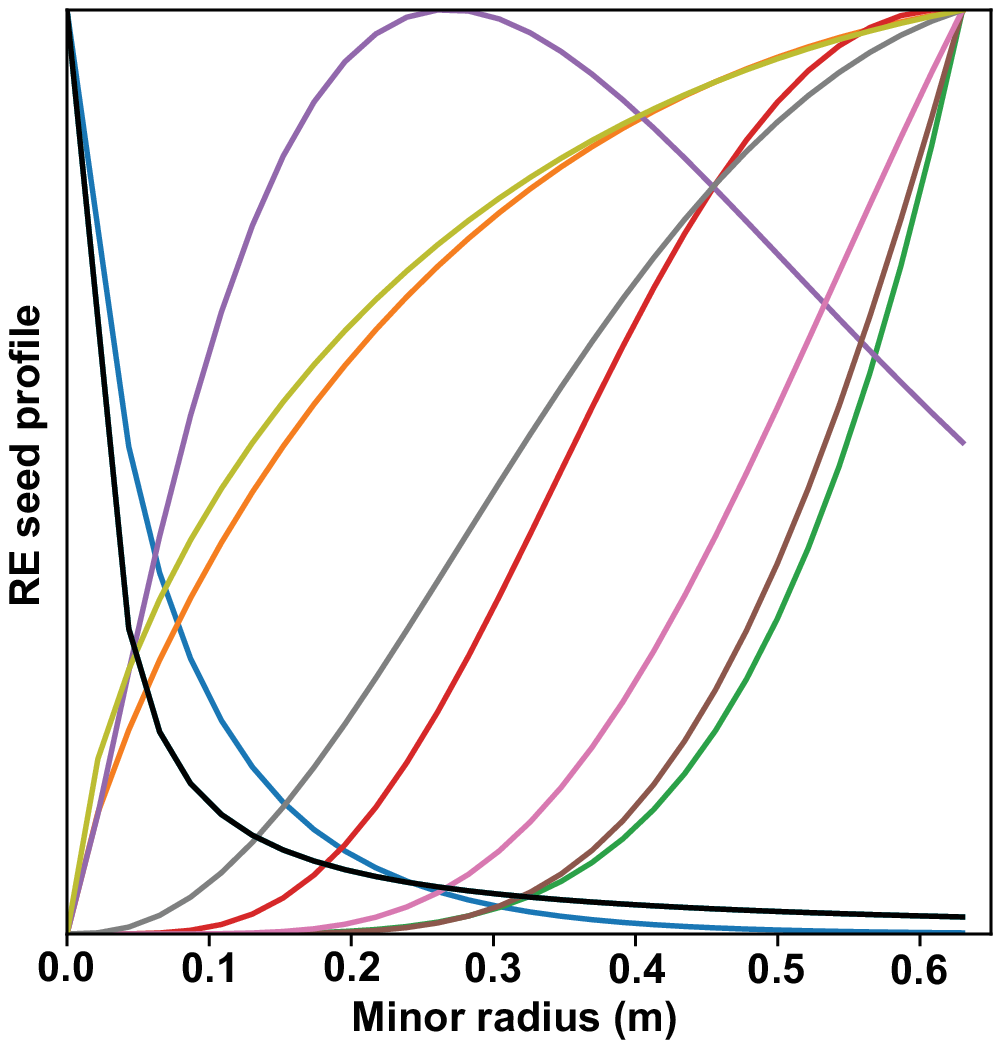}
    \caption{10 randomly sampled RE seed profiles.}
    \label{fig:RE_seed}
\end{figure}

After the 1D proof-of-principle, the next step is to extend the search space to include the other uncertain parameters as well. Since the RE seed is a 1D profile, it is parameterized as the probability distribution function of the gamma distribution: 
\begin{equation}
    f_\text{RE}(r;\alpha,\beta) = \frac{\beta^\alpha r^{\alpha - 1}e^{-\beta r}}{\Gamma(\alpha)},
\end{equation}
where $\alpha$ and $\beta$ are free parameters and $\Gamma(\alpha)$ is the gamma function (Fig. \ref{fig:RE_seed}). The intention is to provide a general parameterization that constrains as little as possible the possible RE seed profile shapes. The search spaces for both $\alpha$ and $\beta$ are set as uniform distributions between 0.001 and 10. 10 randomly sampled RE seed profile shapes are shown in Figure \ref{fig:RE_seed}. It can be clearly seen that this parameterization allows flexible representations of profiles peaking at the center, middle, or edge of the plasma. The search space for the argon assimilation fraction is set as uniform between 0.001\% and 100\%. The search space for the characteristic wall time is set such that $\text{ln}\left(\tau_\text{wall}/1\,\mathrm{ms}\right)$ is sampled uniformly between 0 and 7. As a result, the $\tau_\text{wall}$ values range between about 1 ms and 1100 ms. This allows sampling for very large $\tau_\text{wall}$ parameters exceeding 1.0 s, while encouraging collection of samples at low values. 

Due to the expanded volume of the search space, more samples are expected to be needed than in the 1D example. Therefore, the randmaxvar acquisition function was used with a batch size of 10 samples conducted in parallel. The first 50 samples were collected by sampling the search space randomly, after which the acquisition function was used to recommend sampling locations.

Similar to the 1D search, a rational quadratic kernel is used. The kernel parameters are restricted such that the power parameter in the rational quadratic kernel was constrained to be between $10^{-10}$ and 0.03. The lengthscale constraints are altered for every batch of 10 samples. For even round batches, the lengthscales are constrained to be positive and manually initialized as the distance of the search domain for the dimension divided by the number of collected samples. After this preconditioning step, GP optimization is conducted. There is no maximum lengthscale setup during the even rounds. During the odd round batches, the lengthscales for variations are constrained to be below 1 for the  $T_\text{e}$ and $f_\text{Ar}$ dimensions, below 0.5 for the $\alpha$ and $\beta$ dimensions, and below 0.1 for the $\text{ln}\left(\tau_\text{wall}\right)$ dimension. The lower limits for the lengthscales were set to $10^{-3}$. The even rounds perform essentially automatic relevance determination obtaining very long lengthscales for dimensions that do not show a significant impact on the objective function value, guiding the search to prioritize optimization of input parameters that have more significant impact on the objective function value. However, this approach alone would risk the surrogate model becoming overly confident early in the search and stop exploration for input parameters deemed unimportant. To counteract this risk, the odd rounds apply restrictions of lengthscales, such that the algorithm understands to explore input parameter values, which would simply be extrapolated and interpolated over by the long-lengthscale surrogate model.

\begin{figure}
    \centering
    \includegraphics[width=\textwidth]{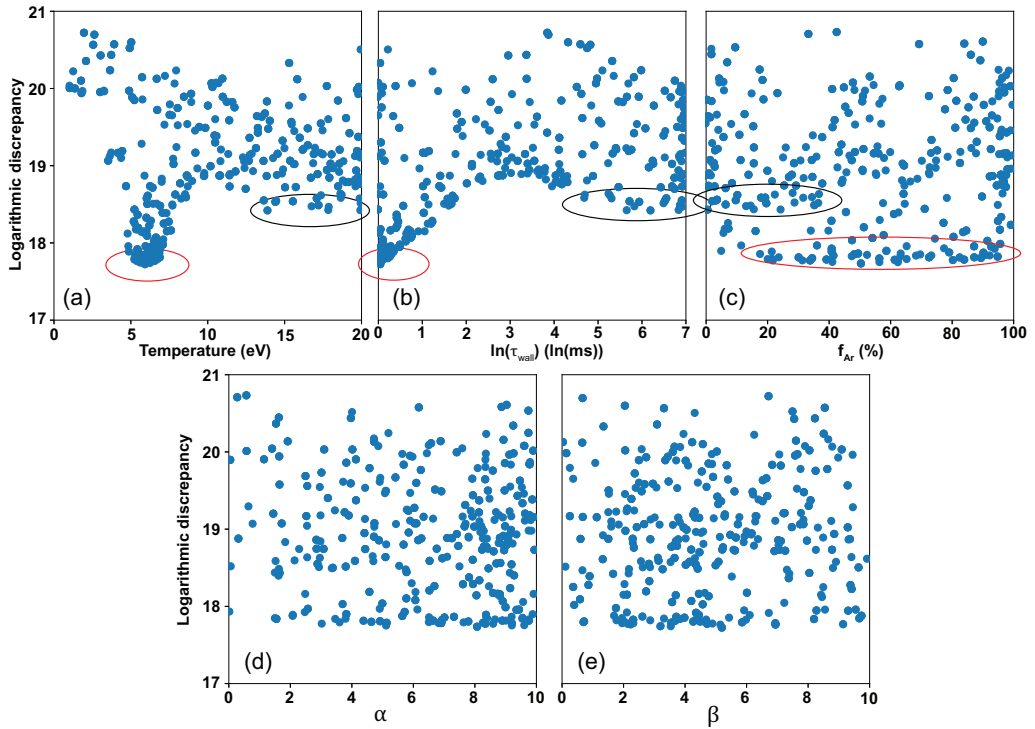}
    \caption{Discrepancies of the collected samples as a function of the dimensions of the search space: (a) temperature, (b) logarithmic characteristic wall time, (c) argon assimilation fraction, and parameters of the runaway seed distribution (d, e). The total number of collected samples is 290. The two optima are highlighted with red (global) and black (local) ellipses.}
    \label{fig:5D_discs}
\end{figure}

The sampling algorithm finds clear optima in the search space for the background plasma temperature and characteristic wall time (Fig. \ref{fig:5D_discs}). It can be observed that there is a global optimum at $T_\text{e}$ around 5 -- 7 eV and $\tau_\text{wall}$ less than about 2.7 ms ($\text{ln}(\tau_\text{wall}/\text{1 ms}) < 1.0$), and a local optimum at $T_\text{e}$ about 15 -- 20 eV and $\tau_\text{wall}$ larger than about 50 ms ($\text{ln}(\tau_\text{wall}/\text{1 ms}) > 4.0$) (Fig. \ref{fig:5D_discs} red and black ellipses). The two solution branches can also be observed in the plot as a function of the argon assimilation fraction, such that the low temperature solution branch reaches the optimum at values above 20 \%, while the higher temperature solution branch at lower values between 5 -- 30 \% (Fig. \ref{fig:5D_discs}c). The shape of the RE seed distribution does not seem to impact the discrepancy significantly (Figs. \ref{fig:5D_discs}d, e).

\begin{figure}
    \includegraphics[width=\textwidth]{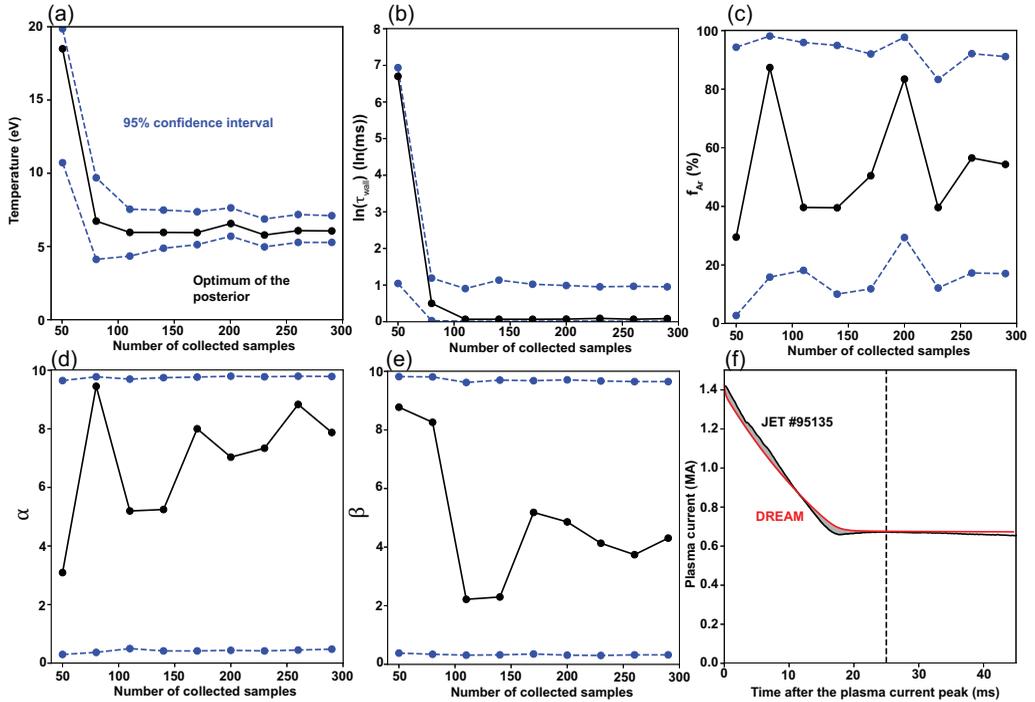}
    \caption{Convergence of the posterior distribution near the global optimum for temperature (a), logarithmic characteristic wall time (b), argon assimilation fraction (c), and RE seed profile parameters $\alpha$ and $\beta$ (d, e) as a function of number of collected samples. The black dots illustrate the current optimum of the posterior distribution and the blue dots with dashed lines the 95 \% confidence interval. (f) The predicted total plasma current (red) with the optimal input parameters after 290 samples compared to the experimental plasma current (black). The GPR in this convergence figure applies the lengthscale restrictions as used for the odd round samples.}
    \label{fig:5D_convergence}
\end{figure}

\begin{figure}
    \centering
    \includegraphics[width=0.8\textwidth]{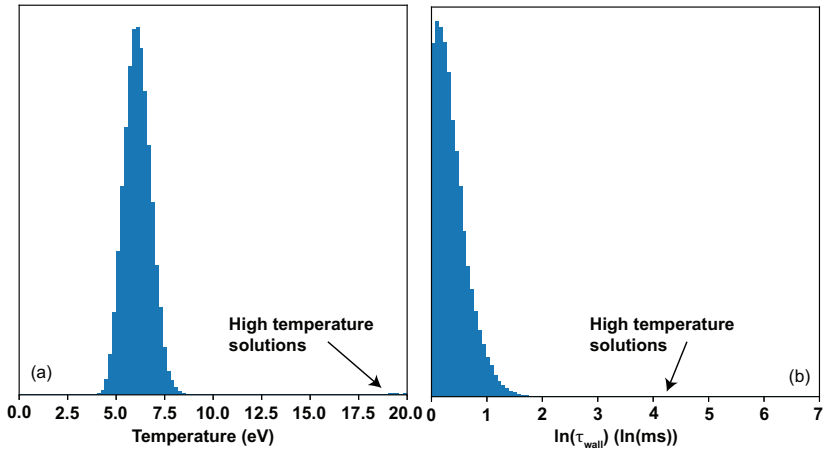}
    \caption{MCMC samples of the posterior distribution for $T_\text{e}$ (a) and $\text{ln} (\tau_\text{wall})$ (b).}
    \label{fig:5D_MCMC}
\end{figure}

After 290 samples, the algorithm estimates the global optimum to be $T_\text{e} \approx 6.1$ eV, $f_\text{Ar} \approx 54.2$ \%, $\tau_\text{wall} \approx 1.1$ ms, $\alpha \approx 7.9$, and $\beta \approx 4.3$, where the 95 \% confidence intervals for $f_\text{Ar}$, $\alpha$, and $\beta$ span most of the search space (Fig. \ref{fig:5D_convergence}). The local uncertainties for the global optimum can be obtained by evaluating the local properties of the approximate posterior. The approximate posterior can be extracted from the probabilistic surrogate model by applying Eq. \ref{posterior}. While a global ABC posterior can be obtained through, for example, MCMC sampling, the local inverse uncertainty near the global optimum is likely of more practical interest in fusion energy research and can be obtained directly from the GPR representation of the posterior. This type of analysis was done by evaluating the one dimensional posterior distribution for each search dimension from the global optimum, which can be integrated to obtain confidence intervals (Fig. \ref{fig:5D_convergence}).  It should be noted that this analysis does not take into account the secondary optimum at higher temperatures as that would require non-linear analysis of the approximate posterior, while MCMC sampling of the posterior would collect some distribution in that area also. However, since multimodality of the optimized function can be observed from the discrepancy plot already (Fig. \ref{fig:5D_discs}), it is considered more important to obtain local uncertainty estimations near the global optimum than sample the full approximate posterior. For completeness, MCMC sampling was conducted for the approximate posterior and the resulting distributions for $T_\text{e}$ and $\text{ln}(\tau_\text{wall})$ show the global optimum, also visible in (Fig. \ref{fig:5D_convergence}), as well as the secondary local optimum at higher temperatures (Fig. \ref{fig:5D_MCMC}).

Analysing the convergence of the search it can be observed that after about 80 -- 110 samples, the posterior distributions for $T_\text{e}$ and $\text{ln}(\tau_\text{wall})$ have converged (Figs. \ref{fig:5D_convergence}a, b). For the other parameters, the 95 \% confidence intervals remain large, indicating that the discrepancy value is not very sensitive to these input parameters, as was observed in the discrepancy plot already (Fig. \ref{fig:5D_discs}). The step observed at 200 samples highlights the stochastic nature of the GPR surrogate model fitting. Depending on the initial conditions, the optimization algorithm might find somewhat different hyperparameters for the GPR, leading to a different shape of the approximate posterior and shift of the optimum and boundaries of the confidence interval as well. However, the large steps of the optimum for $f_\text{Ar}$, as well as the large steps at other sample numbers for $\alpha$, and $\beta$, are a result of the relatively flat posterior distribution shapes, while posterior shape for $T_\text{e}$ and $\text{ln}(\tau_\text{wall})$ are not changed significantly. 

Finally, the predicted and measured total plasma current are compared for the global optimum extracted by the algorithm (Fig. \ref{fig:5D_convergence}f). Comparing the result to the optimum case in the 1D search space example, it can be clearly seen that, with a 5D search space, the algorithm is able to obtain a significantly better fit to the experimentally measured current (Figs. \ref{fig:1D_convergence}c, \ref{fig:5D_convergence}f). In the 1D example, the current was reducing significantly faster than experimentally measured during the early part of the CQ and the end of the CQ happened several ms later than experimentally measured, such that the average L1-norm discrepancy was minimized, while the rate of change of plasma current was poorly matched (Fig. \ref{fig:1D_convergence}c). On the other hand, in the 5D example, the initial drop is also faster than measured experimentally, but soon the rate of change of the plasma current is matched to the experimentally measured rate of change, such that the end of CQ happens near the experimentally measured end of the CQ when the L1-norm between the two currents is minimized (Fig. \ref{fig:5D_convergence}f). The fact that the current is reducing faster than experimentally measured during the early parts of the CQ indicates that the plasma resistance that on average matches the plasma current evolution probably overestimates the resistance during early parts of the CQ. To address this, the final proof-of-principle test conducted in this manuscript is to allow linear variation of $T_\text{e}$ during the simulation by extending the search space to seven dimension.

\subsection{7D search space}\label{7D_search}
As a next extension of the search, a parameterized variation of the background plasma $T_e$ during the simulation is allowed. Linear variation of $T_e$ is assumed, such that the search space is extended to seven dimensions, by adding final temperature, $T_\text{e, final}$, and the time at which $T_\text{e, final}$ is reached, $t_\text{final}$. After $t_\text{final}$ is reached, $T_\text{e}$ is assumed to stay constant. Same search spaces are used for the initial and final $T_\text{e}$, and the search space for  $t_\text{final}$ is set as uniform between 1 ms and 44 ms. 

The rational quadratic kernel is used in the GPR. The power of the kernel was restricted similarly to the setup in the 5D search. Similar to the 5D search, the lengthscale restrictions were altered between even and odd round batches. Batch size was set to 50. For the odd round batches, the lenghscale constraints are similar to those in the 5D search with the same lengthscale contraint applied for the initial and final $T_e$. For the $t_\text{final}$ the minimum lenghtscale is set as 10$^{-3}$ ms and maximum as 1 ms. 

\begin{figure}
    \centering
   \includegraphics[width=\textwidth]{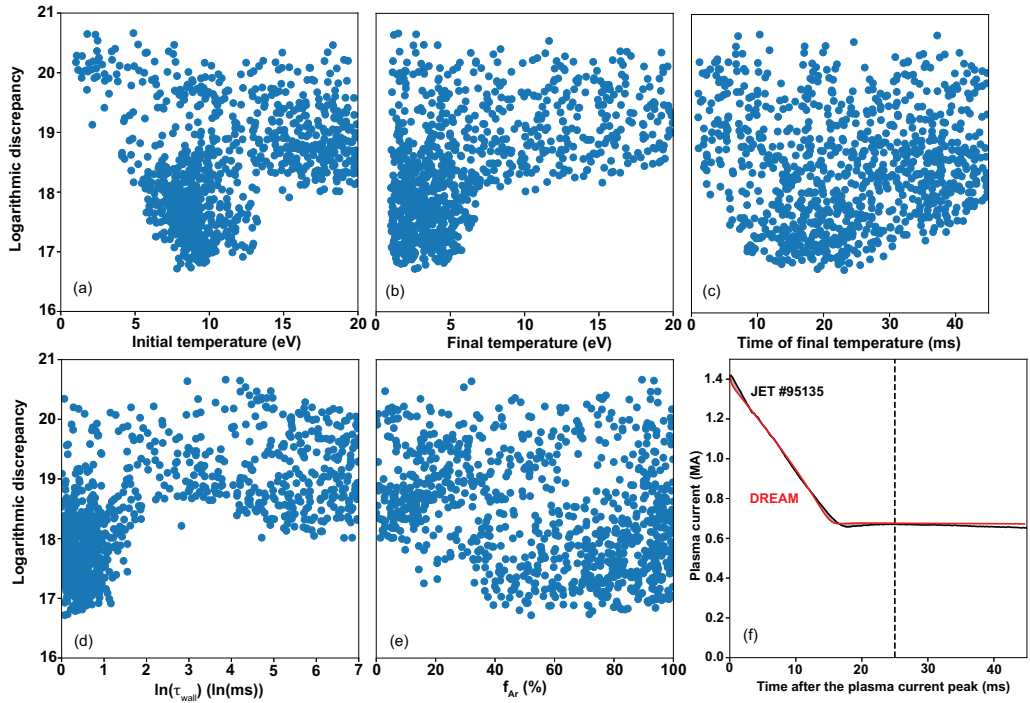}
    \caption{Discrepancies of the collected samples as a function of the dimensions of the search space: (a) Initial temperature, (b) Final temperature, (c) Time at which the final temperature is reached, (d)  logarithmic characteristic wall time, (e) argon assimilation fraction. The $\alpha$ and $\beta$ parameters are not shown as those do not show any significant impact on the discrepancy. (f) The predicted total plasma current (red) with the recommended optimal input parameters compared to the experimental plasma current (black). The total number of collected samples is 950.}
    \label{fig:7D_discs}
\end{figure}

After 950 samples, the local 95 \% confidence intervals around the optimum point, recommended by the algorithm, are $T_\text{e, initial} \in [8.5, 11.7]$ eV, $T_\text{e, final} \in [3.3, 5.8]$ eV, $t_\text{final} \in [11, 19]$ ms, $\tau_\text{wall} \in [1.1, 2.1]$ ms, $f_\text{Ar} \in [48, 97]$ \%, $\alpha \in [0.2, 9.4]$, and $\beta \in [0.5, 9.7]$ (Fig. \ref{fig:7D_discs}). The optimum point is $T_\text{e, initial} \approx 9.5$ eV, $T_\text{e, final} \approx 4.3$ eV, $t_\text{final} \approx 14$ ms, $\tau_\text{wall} \approx 1.4 $ ms, $f_\text{Ar} \approx 76$ \%, $\alpha \approx 4.3$, and $\beta \approx 2.2$. With these input parameters, the predicted current quench rate is very close to the measured values (Fig. \ref{fig:7D_discs}f). Initially, the rate is faster than measured and also the transition to runaway plateau in the simulation occurs about 1 ms earlier than measured (Fig. \ref{fig:7D_discs}f). However, between 3 and 12 ms, the predicted current is nearly exactly on top of the measured current.  

\subsection{Constraining $\tau_\text{wall}$ to 5 ms}\label{4D_6D}

In both the 5D and 7D searches, the algorithm found optimum $\tau_\text{wall}$ around 1.1 - 1.4 ms. This result was somewhat surprising when the conventional prior expectation would have suggested higher values in the range of 5 to 10 ms. Since the optimization algorithm finds the set of parameters that minimize the discrepancy, it is possible that the algorithm compensates for missing physics in the model by reducing $\tau_\text{wall}$ below values that would actually be realistic. As a final test, further 4D and 6D optimizations were conducted, where $\tau_\text{wall}$ was fixed to 5 ms. 

\begin{figure}
    \centering
   \includegraphics[width=\textwidth]{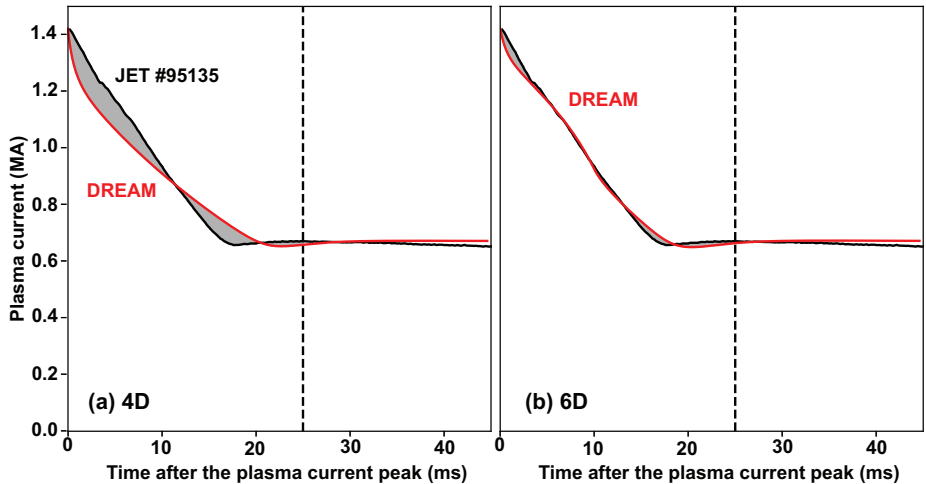}
    \caption{The predicted total plasma current (red) with the recommended optimal input parameters in the 4D (a) and 6D (b) search tasks compared to the experimental plasma current (black).}
    \label{fig:4D_6D_comp}
\end{figure}

In the 4D search, the optimum point recommended by the algorithm is $T_\text{e} \approx 7.0$ eV, $f_\text{Ar} \approx 52$ \%, $\alpha \approx 4.6$, and $\beta \approx 5.5$ (Fig. \ref{fig:4D_6D_comp}a). The local 95 \% confidence intervals are $T_\text{e} \in [6.4, 8.0]$ eV and $f_\text{Ar} \in [32, 91]$ \%, $\alpha \in [0.3, 9.6]$, and $\beta \in [0.3, 9.7]$. In the 6D search, the optimum point recommended by the algorithm is $T_\text{e, initial} \approx 16.2$ eV, $T_\text{e, final} \approx 5.9$ eV, $t_\text{final} \approx 10$ ms, $f_\text{Ar} \approx 82$ \%, $\alpha \approx 1.3$, and $\beta \approx 6.1$ (Fig. \ref{fig:4D_6D_comp}b). The local 95 \% confidence intervals are $T_\text{e, initial} \in [13.9, 19]$ eV, $T_\text{e, final} \in [5.1, 6.7]$ eV, $t_\text{final} \in [8, 12]$ ms, $f_\text{Ar} \in [48, 98]$ \%, and $\alpha \in [0.1, 8.8]$, $\beta \in [0.5, 9.7]$.

As the $f_\text{Ar}$, $\alpha$, and $\beta$ do not impact the discrepancy significantly, the best match obtained by the 4D search seems very similar to the best match obtained by the 1D search (Figs. \ref{fig:1D_convergence}c, \ref{fig:4D_6D_comp}a). When allowing linearly varying background plasma $T_\text{e}$, the algorithm is able to find a solution that matches the experimentally measured plasma current nearly as well as in the 7D search (Figs. \ref{fig:7D_discs}f, \ref{fig:4D_6D_comp}b). However, when fixing $\tau_\text{wall} = 5$ ms, the recommended initial $T_\text{e}$ is increased from 9.5 to 16.2 eV, and $t_\text{final}$ reduced from 14 ms to 10 ms, highlighting the non-linear dependencies between the optimal input parameters.   

\subsection{Convergence as a function of dimensions}\label{conv}

\begin{figure}
    \centering
   \includegraphics[width=0.9\textwidth]{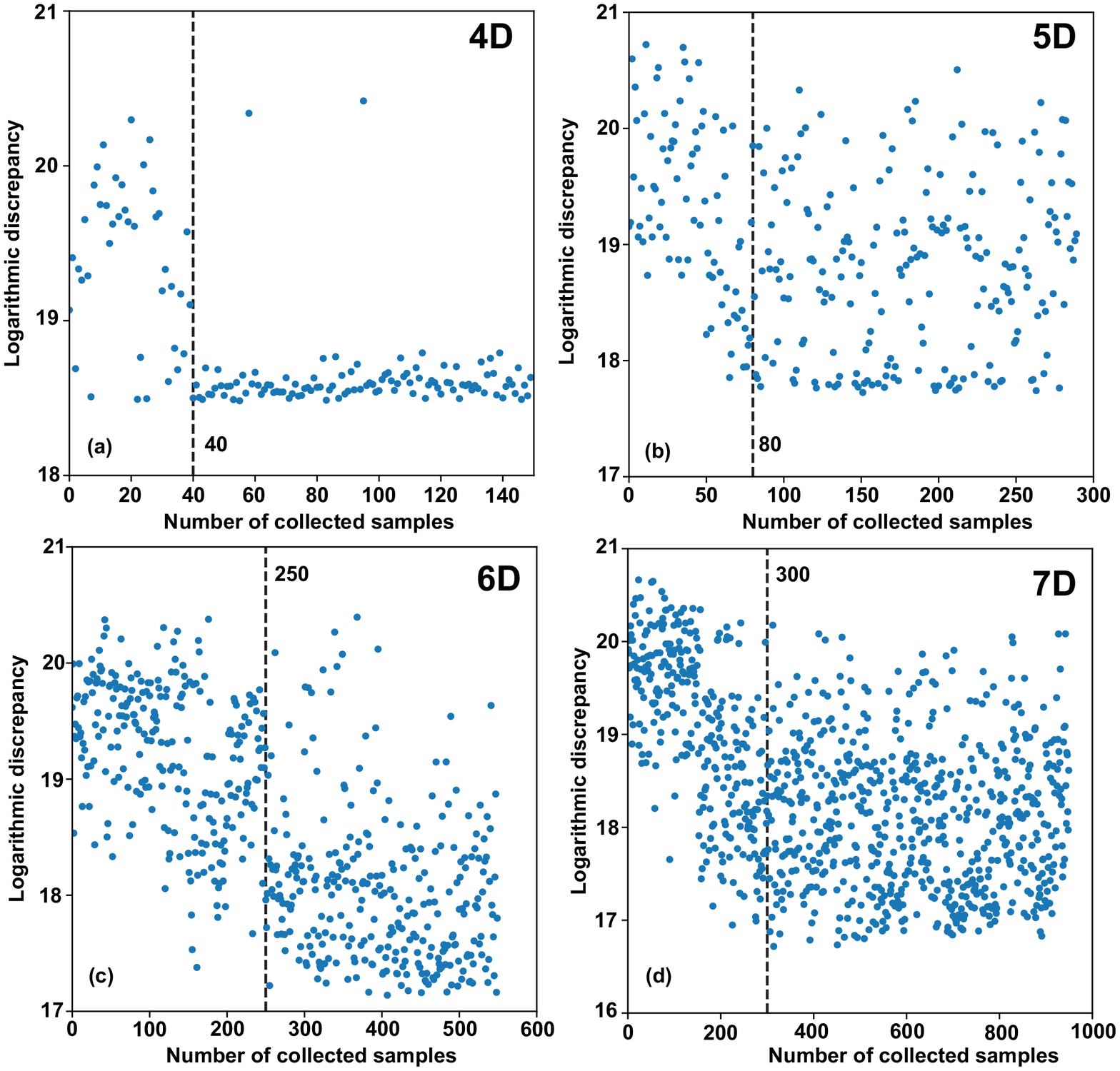}
    \caption{Discrepancy as a function of sample number for the 4D (a), 5D (b), 6D (c), and 7D (d) search tasks. The vertical dashed lines illustrate the approximate point when the minimum discrepancy saturates.}
    \label{fig:conv_per_sample}
\end{figure}

The computational challenge of the optimization task increases with the number of dimensions of the search space. Figure \ref{fig:conv_per_sample} illustrates the discrepancy as a function of sample numbers for the 4D, 5D, 6D, and 7D search tasks in this work. Evaluating the convergence based on the sample number after which the minimum discrepancy saturates, the 4D search converges after about 40 samples, the 5D search after about 80 samples, the 6D search after about 250 samples, and the 7D after about 300 samples. Beyond this point, increasing sample numbers will reduce the uncertainty of the posterior distribution while the minimum discrepancy is not reduced anymore. 

\begin{figure}
    \centering
   \includegraphics[width=0.7\textwidth]{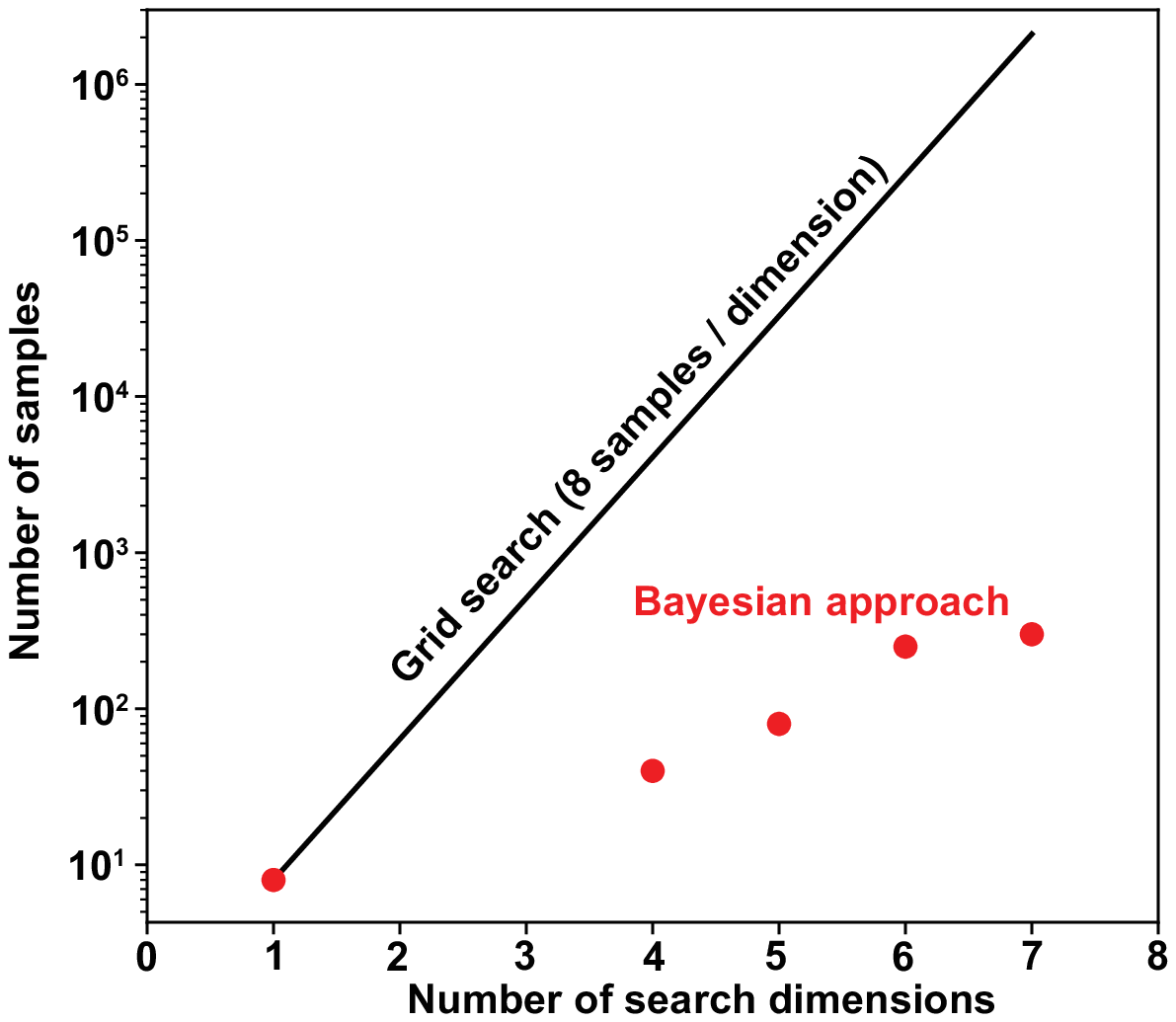}
    \caption{Number of samples as a function of number of search dimensions: grid search with 8 samples for each dimension (black line), Bayesian approach in this work (red circles).}
    \label{fig:GS_vs_BO}
\end{figure}

Comparing to a grid search of eight samples for each dimension, the Bayesian optimization algorithm is very efficient (Fig. \ref{fig:GS_vs_BO}). Beyond four dimensions, the grid search algorithm would be calling for over 10000 samples and soon become intractable. The Bayesian approach, on the other hand, obtains samples near the minimum discrepancy after a few hundred samples even in the case of the seven dimensional search space.

\section{Summary}

Bayesian approach has been explored for validation of runaway electron simulations. Many of the simulation tools
applied in fusion energy research require the user to specify several input parameters that are not constrained by the available experimental information. Bayesian
inference algorithms offer a promising approach to determine these free parameters with uncertainty quantification and is less subjective to user bias than approaches based on manual parameter calibration. The main challenge in using an algorithmic approach to parameter calibration is the computational cost of
simulating enough samples to construct the posterior distributions for the uncertain
input parameters. By using probabilistic surrogate modelling, through Gaussian Process regression, with Bayesian optimization, it is possible to reduce the number of required simulations by several orders of magnitude.  This type of Bayesian optimization framework was implemented in this work for a disruption runaway
electron analysis model, and explored for current quench simulations for a JET plasma
discharge with an argon induced disruption. The algorithm is able to find optimal input parameters with uncertainties in one to seven dimensional proof-of-principle cases, and is several orders of magnitude more sample efficient than a regimented grid search algorithm would have been. 

Surrogate model specification is central to the performance of the search algorithm. Using the Gaussian process approach, the kernel parameters need to be appropriately constrained for the surrogate model to provide meaningful guidance for the search through the acquisition function. 
An overly smooth kernel with long maximum correlation lengthscales can make the surrogate model overly confident and not find the actual global optimum. On the other hand, limiting the lengthscales to small values will encourage exploration but also require more iterations for convergence. Finding the appropriate surrogate model specifications is an area that requires attention from the user of these algorithms. The most appropriate constraints are likely to be specific to each search task. Furthermore, both specifying appropriate kernel constraints and diagnosing potential issues with kernel constraints become more challenging with increasing number of search dimensions. Therefore, it would be desired to find default kernel constraints that are likely to work acceptably well in most circumstances. The approach chosen here was to alternate the kernel constraints at specific sample intervals between unconstrained but positive lengthscales and lengthscales constrained to be below a certain threshold that is a fraction of the width of the search dimension. By alternating the kernel constraints, the risk of the algorithm either oversmoothing a region or getting into a mode of infinite exploration is reduced. However, more generally applicable methods for constraining the surrogate model are likely to exist, and could improve the performance of the algorithm further.

\section*{Acknowledgements}
This work has been carried out within the framework of the EUROfusion Consortium, funded by the European Union via the Euratom Research and Training Programme (Grant Agreement No 101052200 — EUROfusion). Views and opinions expressed are however those of the authors only and do not necessarily reflect those of the European Union or the European Commission. Neither the European Union nor the European Commission can be held responsible for them. The work of Eero Hirvijoki was supported by the Academy of Finland grant no. 315278. The work was supported in part
by the Swedish Research Council (Dnr.~2018-03911). The authors wish to acknowledge CSC – IT Center for Science, Finland, for computational resources.

\bibliographystyle{jpp}
\bibliography{mybib}

\end{document}